\documentclass[sn-nature]{sn-jnl}

\usepackage{graphicx}%
\usepackage{multirow}%
\usepackage{amsmath,amssymb,amsfonts}%
\usepackage{amsthm}%
\usepackage{mathrsfs}%
\usepackage[title]{appendix}%
\usepackage{xcolor}%
\usepackage{textcomp}%
\usepackage{manyfoot}%
\usepackage{booktabs}%
\usepackage{algorithm}%
\usepackage{algorithmicx}%
\usepackage{algpseudocode}%
\usepackage{listings}%
\usepackage{wasysym}
\usepackage{textcomp}
\usepackage{comment}
\usepackage{amssymb}
\usepackage{ bbold }

\theoremstyle{thmstyleone}%
\theoremstyle{thmstyletwo}%

\theoremstyle{thmstylethree}%

\begin{document}

\title[Article Title]{Quantitative phase imaging via a multimode fiber}


\author[1,2]{\fnm{Aleksandra} \sur{Ivanina}}
\equalcont{These authors contributed equally to this work.}

\author[1,3]{\fnm{Maxim} \sur{Marshall}} 
\equalcont{These authors contributed equally to this work.}

\author[1]{\fnm{Ksenia} \sur{Abrashitova}}

\author[4]{\fnm{Tristan} \spfx{van} \sur{Leeuwen}}

\author*[1,2]{\fnm{Lyubov} \spfx{V.} \sur{Amitonova}}\email{l.amitonova@vu.nl}

\affil*[1]{\orgname{Advanced Research Center for Nanolithography (ARCNL)}, \orgaddress{\street{Science Park 106}, \city{Amsterdam}, \postcode{1098 XG}, \country{The Netherlands}}}

\affil[2]{\orgdiv{Department of Physics and Astronomy}, \orgname{Vrije Universiteit (VU) Amsterdam}, \orgaddress{\street{De Boelelaan 1081},  \postcode{1081 HV}, \country{The Netherlands}}}

\affil[3]{\orgname{University of Amsterdam}, \orgaddress{\street{Science Park 908},  \postcode{1098 XH}, \country{The Netherlands}}}

\affil[4]{\orgname{Centrum Wiskunde \& Informatica (CWI)}, \orgaddress{\street{Science Park 123},  \postcode{1098 XG}, \country{The Netherlands}}}


\abstract{Label-free quantitative phase imaging is a vital tool for optical microscopy and metrology applications. A hair-thin multimode fiber stands out as a very attractive platform for minimally invasive imaging. Here we propose and experimentally demonstrate a non-interferometric non-iterative approach for high-speed high-resolution label-free quantitative phase imaging via a multimode fiber, unlocking multiple applications in life science and bioimaging.}


\maketitle

\section{Introduction}\label{sec1}
Multimode fibers (MMFs) have recently emerged as an ultimate endoscopic technology that enables high-resolution imaging at the tip of a hair-thin flexible probe \cite{cao2023controlling}. A wide range of imaging modalities through MMF-based endoscopes have been demonstrated, including (auto-)fluorescence \cite{vcivzmar2011shaping, ohayon2018minimally, lochocki_epi-fluorescence_2022, stiburek_110_2023}, confocal reflectance~\cite{loterie_confocal_2015, lee_confocal_2022}, chemically-selective \cite{saar2011coherent,tragaardh2019label}, super-resolution \cite{amitonova2020endo,abrashitova2022high}, and 3D \cite{stellinga2021time, pascucci2019compressive} imaging. However, even nowadays, state-of-the-art fiber-based microscopy relies on amplitude-only contrasts. While to visualize `phase objects' and to measure many morphological parameters of tissue, such as refractive index variations or matter density, quantitative phase information is required~\cite{nguyen2022quantitative, park2018quantitative, hillmann_vivo_2016}.
As conventional detectors record only the amplitude of light, advanced approaches are needed to recover phase information. This fundamental problem stimulates the development of many techniques.

Quantitative phase imaging (QPI) is a set of label-free methods to measure the optical path length delays within the sample.
The problem of recovering the complex wavefront can be solved by interferometric approaches that, unfortunately, require extreme environmental stability and a high-quality reference beam~\cite{zhang2021review}.
A broad range of techniques that eliminates the use of a reference include Fourier ptychography~\cite{zheng2021concept}, coherent diffractive imaging~\cite{gerchberg1972practical}, and differential phase contrast microscopy~\cite{tian_3d_2014}. These methods reconstruct the phase shift from a relatively large sequence of intensity patterns and require iterative computationally expensive post-processing.
One of the most popular non-interferometric QPI methods is Shack–Hartmann (SH) wavefront sensing~\cite{platt2001history}. A lenslet array creates focused spots, the position of which depends linearly on the local phasefront. However, the lenslet manufacturing parameters -- total number, focal length, curvature, and diameter of lenses -- limit the spatial resolution and sensibility. Multiple solutions that utilize structurally modulated light have been proposed to overcome the limitations of SH sensor~\cite{zhang_phase_2007, gao_phase_2013, chowdhury_structured_2014, soldevila_phase_2018}. However, they impose strict restrictions on the illumination projection system.

Despite all the recent progress, the development of a compact high-resolution endoscopic QPI remains an important goal and a big challenge.
Recently, QPI through a multicore fiber has been demonstrated~\cite{sun2022quantitative}. However, it requires a complex probe with knowledge of the full transmission matrix, capturing four high-resolution images, and subsequent iterative reconstruction.
High spatial resolution is another major challenge of QPI, especially for compact non interferometric sensors~\cite{cotte_marker-free_2013, mico_resolution_2019}. The lateral resolution of optical imaging is generally limited by the diffraction of light:
\begin{equation}
\mu_{min} = k\frac{\lambda}{NA}, 
\label{eq:limit}
\end{equation}
where $\mu_{min}$ is the smallest resolvable distance, $\lambda$ is the wavelength of the incident radiation and NA is the numerical aperture of the optical system. The critical parameter $k$ is determined by the imaging modality and yields either $0.5$ or $0.82$ for conventional incoherent and coherent imaging, respectively. Therefore, the lateral resolution of  QPI is generally determined by the $0.82\lambda/NA$ diffraction limit~\cite{mico_resolution_2019,born2013principles}.

Here, we propose and experimentally demonstrate non interferometric quantitative phase imaging via a hair-thin MMF. An appealing feature of the reconstructed complex amplitude of a wavefront is that the lateral resolution is limited by the incoherent diffraction limit of the illumination, which is $1.5$ times better than the diffraction limit of conventional QPI. Moreover, we eliminate the need for a high-resolution camera and rely solely on a position-sensitive detector (PSD). With a high-speed PSD, the imaging speed is limited by the scanning rate and the total number of measurements and can reach $50$ frames per second with the present hardware. 
We experimentally demonstrate quantitative phase reconstruction of a variety of phase targets with a simple noniterative algorithm. Our approach paves the way toward ultracompact phase sensors with applications in biology, medicine, and the semiconductor industry.

\section{Main principle}\label{sec2}

\begin{figure}[t]%
\centering
\includegraphics[width=0.95\textwidth]{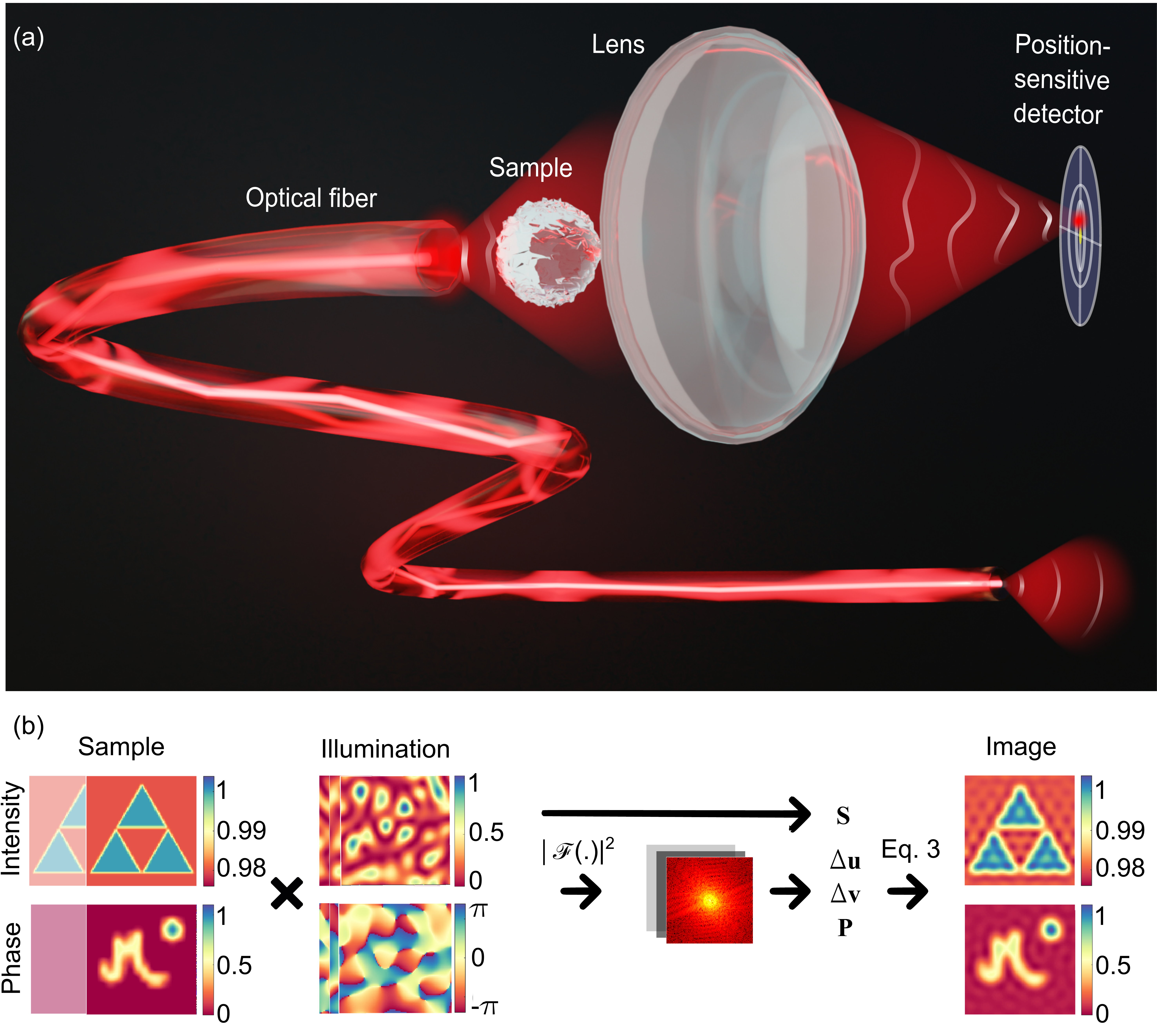}
\caption{Illustrations of the proposed MMF-based QPI. (a) Schematic of the experimental setup.
(b)~Experimental workflow depicted through the simulation results. Complex speckle fields generated by an MMF (second column) are used to illuminate the unknown sample (first column). The center of mass coordinates in the Fourier plane with respect to the `reference' sample, $\Delta u_n, \Delta v_n$, and the total transmitted power, $P_m$, are recorded by the position-sensitive detector. These measurements together with matrix $\bold{S}$ of the flattened intensity distributions of the speckle patterns are used to reconstruct the sample phase and amplitude distributions.}
\label{idea}
\end{figure}

The main idea of the proposed approach is illustrated in Fig.~\ref{idea}(a). Coherent light coupled into an MMF gives rise to a complex speckle pattern, $B_m(x,y) e^{i\psi_m(x,y)}$.
We use a set of these seemingly random patterns to sample the unknown wavefront, $A(x,y)e^{i\phi(x,y)}$.
Each pattern illuminates the sample and a single lens is used to form phase-sensitive intensity distributions in the Fourier plane -- a principle also fundamental to Shack–Hartmann sensors. A position-sensitive detector in the back focal plane of the lens records the coordinates of the center of mass, $(u_m, v_m)$ and the total transmitted intensity, $P_m$.
Both, the amplitude and phase of the sample can be computationally recovered using the linear relationship between the local wavefront slope at the sample plane and centroid displacement in a Fourier plane~\cite{goodman_introduction_2005, Ghiglia82}.
As shown in Supplementary 1.1-1.2, the position of the centroid ($u_m$, $v_m$) is linked to the gradient of the phase by:

\begin{align}
		\label{eqn:main2}
		\begin{split}
			\begin{pmatrix} u_m \\ v_m \end{pmatrix} = \frac{\lambda \text{F}}{2\pi} \frac{1}{P_m}\iint (B_{m}(x,y) A(x,y))^2 \,\nabla \left( \phi(x,y) + \psi_m(x,y) \right)  \, dx \, dy,
		\end{split}
	\end{align}
where F is the focal distance of the lens; $A(x,y) \le 1$ is the amplitude attenuation of the sample; $\phi(x,y)$ is the phase delay introduced by the sample; $B_m(x,y)$ is the amplitude and $\psi_m(x,y)$ is the phase of the $m$-th speckle pattern; $P_{m} = \iint B_m^2(x,y)A^2(x,y)\, dx \, dy$ is the total power transmitted through the sample for each speckle pattern and $\nabla = (\partial/\partial_x, \partial/\partial_y)$ is the gradient in two dimensions.

The main challenges of solving Eq.~\ref{eqn:main2} for unknown $\phi(x, y)$ and $A(x,y)$ include the computational reconstruction of the wavefront at all points in space from `single-pixel' measurements (centroid position), demultiplexing, and the distillation of the sample field from the field of complex illumination patterns.
As shown in Supplementary 1.6, Eq.~S23, the problem can be solved directly if the full calibration of the imaging system is performed and both amplitude $B_m(x,y)$ and phase $\psi_m(x,y)$ of illuminating speckle patterns are known.

While this full transmission matrix pre-calibration is experimentally possible \cite{popoff_measuring_2010}, in real-life scenarios it might be difficult to measure phase profiles of the speckle patterns.
We propose an elegant solution to reconstruct the unknown sample wavefront using only intensity maps $S_m(x,y)=B^2_{m}(x,y)$ of the speckle patterns. This approach works for phase-only samples as well as for measuring (e.g. time-dependent) phase changes of a sample with a constant amplitude profile. The key is to implement a differential measurement scheme and use a set of centroid positions ($u_{m0}$, $v_{m0}$) of a reference, e.g. flat phase, sample. For continuous monitoring of phase changes of a given sample, ($u_{m0}$, $v_{m0}$) are the coordinates of the centroid position at a previous time-stamp. In these cases, the gradient of the sample phase can be reconstructed without knowing the complex phase distributions of illuminating speckle patterns, as derived in Supplementary 1.3-1.5. 

We now represent the unknown amplitude and phase by vectors 
$\mathbf{a}\in\mathbb{R}^N$, $\partial\phi/\partial_x \equiv \boldsymbol{\phi}_x \in \mathbb{R}^N,  \partial\phi/\partial_y \equiv \boldsymbol{\phi}_y\in\mathbb{R}^N$ which represent the squared amplitudes $A(x,y)^2$ and phase gradient $\nabla\phi(x,y)$ on a grid of $N$ pixels. Similarly, we represent the measured relative centroid positions in the $x$ and $y$ directions corresponding to the $M$ speckle patterns by $\Delta\bold{u}\in\mathbb{R}^M$, $\Delta\bold{v}\in\mathbb{R}^M$.

We now reconstruct the unknown amplitude and phase from these measurements as follows:
\begin{equation}
\begin{aligned}
\mathbf{a} &= \mathbf{S}^{\dagger}\mathbf{p},\\
\left(\begin{matrix} \boldsymbol{\phi}_x \\ \boldsymbol{\phi}_y\end{matrix}\right)&=\left(\begin{matrix}\bold{M}^{\dagger} & \bold{0} \\ \bold{0} & \bold{M}^{\dagger}\end{matrix}\right)\left(\Delta\bold{u}\atop{\Delta\bold{v}}\right),
\label{solutioneq}
\end{aligned}
\end{equation}
where
\begin{equation}
\bold{M} = \frac{\lambda \text{F}} {2\pi}\operatorname{diag}(\textbf{p})^{-1}\textbf{S}\operatorname{diag}(\textbf{a}),
\end{equation}
with $\mathbf{S}\in\mathbb{R}^{M\times N}$ consisting of $M$ rows of $\operatorname{vec}(B_m(x,y)^2)$, and
$\mathbf{p}\in\mathbb{R}^M$ a vector with the total transmitted power, $P_m$, for each speckle pattern.
The operator $\text{diag}(\cdot)$ constructs a matrix with the given vector on its diagonal, the operator $\text{vec}(\cdot)$ converts the matrix into a vector, and $\cdot^\dagger$ denotes the Moore-Penrose pseudo-inverse~\cite{gong2015high}.
The amplitude map $A(x,y)$ is reconstructed by reshaping and calculating the square root of $\mathbf{a}$ and the quantitative phase map $\phi(x,y)$ is calculated by reshaping and numerically integrating $\boldsymbol{\phi}_x, \boldsymbol{\phi}_y$.

The experimental workflow is schematically shown in Fig.~\ref{idea}(b) where we also present our simulation results. We simulate the whole procedure of the proposed MMF-based QPI for a complex sample ($92\times 92$ pixels) presented in Fig.~\ref{idea}(b, first column) together with a `reference' wavefront. As an illumination, a set of $92\times 92 = 8464$ fully-developed random speckle patterns with the diffraction-limited speckle size is generated~\cite{goodman2007speckle}.  Examples of illumination  phase $\psi_{m}(x,y)$ and intensity $S_{m}(x,y)$  distributions are presented in Fig.~\ref{idea}(b, second column). Each speckle pattern is multiplied by a sample complex transmission function $A(x,y)e^{\textit{\textbf{i}}\phi(x,y)}$. The intensity distributions in the back focal plane of the lens are simulated by the Fourier transform~\cite{goodman_introduction_2005}. The total transmitted power and the centroid coordinates for each speckle illumination pattern have been calculated. The procedure is repeated for the reference object with a plane phase profile.
The complex sample wavefront is reconstructed by Eq.~\ref{solutioneq} as presented in Fig.~\ref{idea}(b, right).
The details of the procedure can be found in Supplementary 1.6.
The reconstructed phase and amplitude are in excellent agreement with the original sample.
Our simulations confirm that the proposed workflow can be used for MMF-based QPI.
 
\section{Results}\label{results} 
\begin{figure}[t]%
\centering
\includegraphics[width=0.95\textwidth]{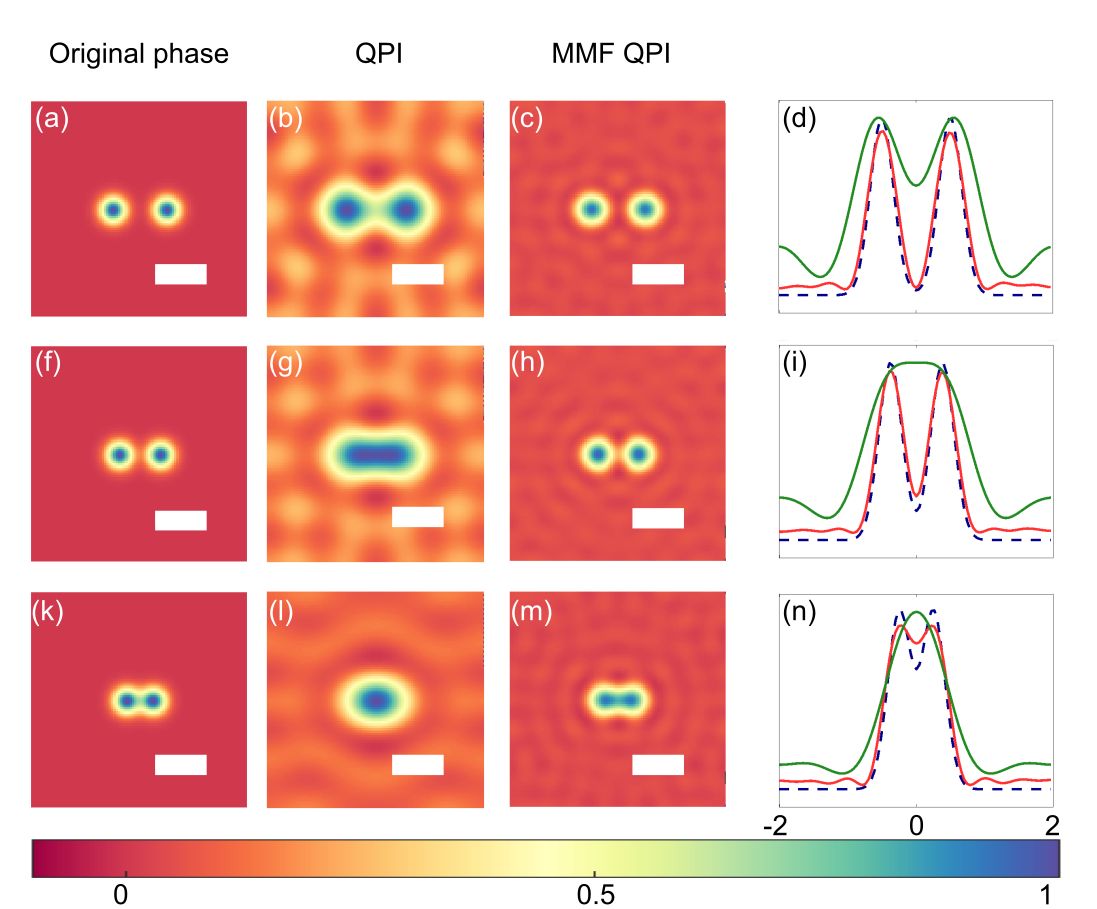}
\caption{MMF QPI simulation results. (a,f,k) Phase samples with different distances between the dots:  $\lambda/NA$ (a), $0.76\lambda/NA$ (f), and $0.5\lambda/NA$ (k). (b,c,g,h,l,m) Phase imaging results for conventional coherent imaging (b,g,l) and proposed MMF imaging (c,h,m).
(d,i,n) Horizontal cross-sections along the peaks for the ground-truth phase (blue dashed lines), conventional coherent imaging (green solid lines), and MMF imaging (red solid lines). Scale bars are coherent diffraction limit $0.82\lambda/NA$.
}
\label{fig5}
\end{figure}

The proposed MMF-based QPI can be seen as a generalized ultra-compact version of a SH phase sensor. The spatial resolution of the SH type of sensors is limited by the lenslet size~\cite{wang2017ultra} or a pattern aperture size \cite{soldevila_phase_2018},  which in our case is equal to the average speckle size of $0.5\lambda/NA$. Therefore, the resolution of MMF-based QPI is expected to be higher than the resolution of conventional, e.g. holographic, coherent phase imaging techniques, which are limited by $0.82\lambda/NA$.
To characterize the lateral resolution of the proposed approach, we simulate imaging of phase-only samples consisting of two phase objects separated by different distances: $\lambda/NA$, $0.76\lambda/NA$, and $0.5\lambda/NA$, as depicted in Figure~\ref{fig5}(a,f,k), respectively.
To simulate conventional coherent diffraction-limited imaging without the loss of generality, we assume that the phase sample is illuminated with a plane wave and account for diffraction effects by applying the low-pass circular binary filter with the cutoff frequency $\text{NA}/\lambda$ in the Fourier plane, as discussed in Supplementary 2. The MMF-based QPI has been simulated as described in the Methods section (see Fig.~\ref{idea}) using a set of $92\times 92 = 8464$ fully-developed random speckle patterns with the diffraction-limited speckle size.

The results are presented in Fig.~\ref{fig5}(b,g,l) for conventional phase imaging and in Fig.~\ref{fig5}(c,h,m) for the proposed MMF-based QPI. The horizontal cross-sections along the centers of peaks are shown in Fig.~\ref{fig5}(d,i,n) by the solid green line for the conventional and by the solid red line for the proposed approach, respectively. The ground truth is shown by the dashed blue line.
As expected (Eq.~\ref{eq:limit} for the coherent case), the phase structures are distinguishable in Fig.~\ref{fig5}(b) and not resolved in Fig.~\ref{fig5}(g) and (l), where the feature size is smaller than $0.82\lambda/NA$. In contrast, the proposed speckle-based QPI allows to resolve peaks at shorter distances as can be seen in Figure~\ref{fig5}(h, m) up to a limit of $0.5\lambda/NA$. Our simulations confirm that the lateral spatial resolution of the proposed MMF QPI is limited by the speckle size, which is $0.5\lambda/NA$,  providing a resolution enhancement of approximately $1.5$ times compared to conventional coherent diffraction-limited phase imaging.

\begin{figure}[h!]%
\centering
\includegraphics[width=0.9\textwidth]{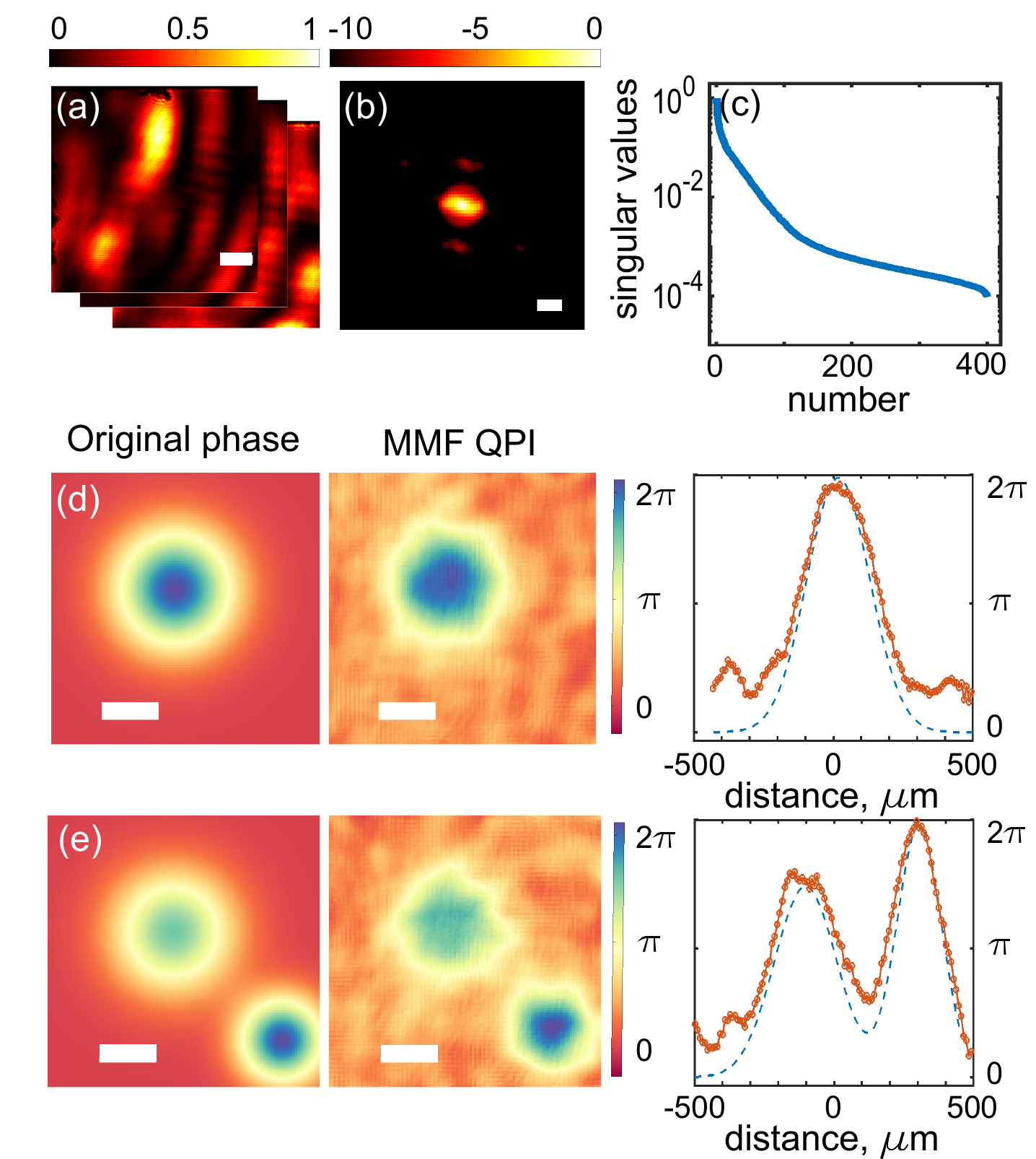}
\caption{Experimental results on MMF QPI.
(a) Speckle patterns generated by the MMF, $S_m(x,y)$. The scale bar is $0.5\lambda/\text{NA}\approx92$~$\mu$m. (b) Zoomed-in average power spectrum density of speckle patterns. The scale bar is $2\text{NA}/\lambda$. (c) Singular value decomposition of \textbf{S}. (d,e)
MMF-based phase imaging results: phase patterns projected on the SLM (left), reconstructed phase images (middle) and the cross-sections along the center of the peaks (right) of the original patterns (the dashed blue line) and their corresponding images (the orange line). The scale bars are $0.82\lambda/NA = 150~\mu$m. }
\label{fig3}
\end{figure}

We experimentally demonstrate the MMF-based QPI using a spatial light modulator (SLM, Meadowlark, $1920\times1200$, pxl size $8~\mu$m) as a versatile re-configurable phase sample. The detailed experimental setup is presented in Fig.~S5 and fully described in Supplementary 3. The simplified drawing is shown in Fig.~\ref{idea}(a). The input facet of a standard MMF with NA$_{\text{fiber}}$ = 0.22 and a diameter of $50~\mu$m is illuminated by a focused coherent linearly polarized laser beam with a wavelength of $\lambda$~=~640~nm.
A set of $400$ different speckle illumination patterns at the output fiber facet is generated by scanning the focal spot at the input on a uniform grid. The patterns are magnified 63$\times$, such that the illumination NA is NA$_{\text{fiber}}/63 = 3.5\cdot10^{-3}$,  and then projected onto the sample (SLM screen). The field of view is restricted to 780~$\mu$m~by~780~$\mu$m or 90$\times$90 SLM pixels using a square pinhole in an intermediate image plane.
The intensity distributions of the speckle patterns, $S_m(x,y)$, are recorded by a camera during the pre-calibration step. The examples of illumination patterns are presented in Fig.~\ref{fig3}(a). By investigating the averaged frequency content of the patterns, as presented in Fig.~\ref{fig3}(b), we confirm that the average speckle intensity size is $0.5\lambda/NA$.
The diffraction limit together with the limited field-of-view put a threshold on the total number of uncorrelated speckle patterns. In our experimental configuration, the number of independent speckle patterns is estimated to be about $100$. It matches with the rank of the experimentally measured matrix $\bold{S}$, as presented in Fig.~\ref{fig3}(c).
We align the setup by imaging 2D linear phase gradients as discussed in Supplementary 3.3.

\begin{figure}[t]%
\centering
\includegraphics[width=1\textwidth]{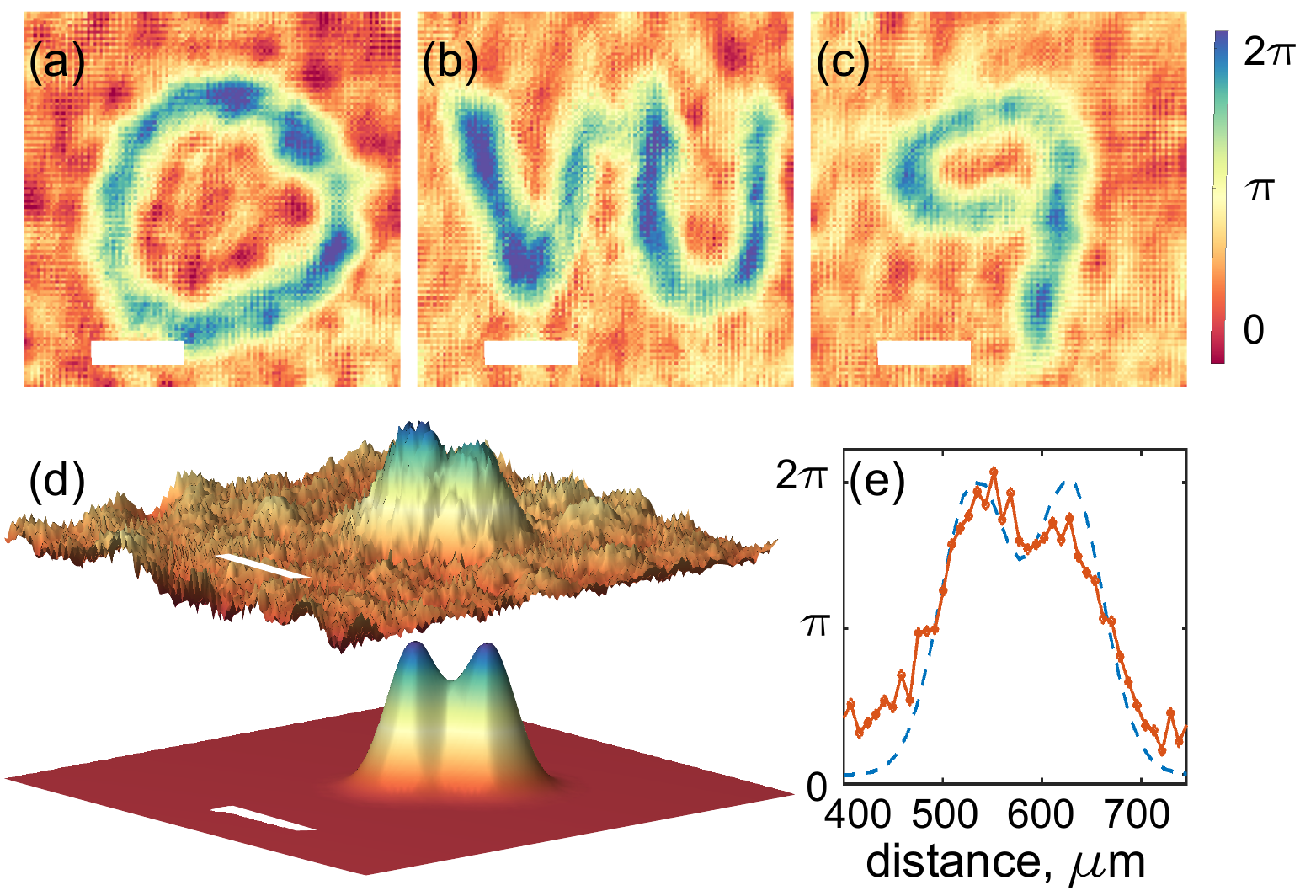}
\caption{MMF-based QPI of complex phase targets. (a-c) Experimentally measured phase patterns representing letter `O' (a), university logo `VU' (b), and digit `9' (c). (d) Experimental demonstration of high spatial resolution: (d, top) Experimentally measured image of a phase pattern consisting of two Gaussian profiles at a close distance of $92~\mu$m, which matches the theoretical resolution limit of the MMF-based QPI, $0.5\lambda$/NA. (d, bottom) The original SLM pattern.
(e) Cross-section along the centers of Gaussian peaks for the SLM pattern (the dashed blue line), and MMF-based image (the orange line). The scale bars are $0.82\lambda/NA = 150~\mu$m.
}
\label{fig4}
\end{figure}

In the first set of experiments, we image different Gaussian phase distributions projected on the SLM. As an illumination, $200$ random speckle patterns on the output of an MMF have been used.
The results are presented in Fig.~\ref{fig3}(d,e). 
We see an excellent quantitative match between the reconstructed images and the original phase patterns. The cross-section along the centers of Gaussian
peaks of the original patterns (dashed blue lines) and phase images (orange lines) are in very good agreement. 
The experimentally measured images contain a grid-like structure and an oscillatory background, which can be explained by several main factors, apart from evident experimental noise. The reconstruction algorithm uses hard thresholding of singular values, which can be seen as a generalization of Fourier low-pass filtering. This kind of background is fundamental to coherent imaging techniques and can also be seen in simulated noise-free images both for conventional and proposed approaches (Fig.~\ref{fig5}). In addition, matrix $\bold S$ is recorded in reflection from the SLM, which results in the superposition of the SLM pixels with the incident speckle patterns. The periodic pixel boundaries are becoming visible in the final images. The details can be found in supplementary 2.1 and 2.2.

In the next set of experiments, we image different complex phase distributions using $400$ speckle patterns generated in a MMF. The results for letter `O', university logo `VU' and digit `9' are presented in Fig.~\ref{fig4}(a-c). 

Finally, to experimentally investigate the resolution limits of the proposed MMF QPI, we image the sample consisting of two Gaussian phase profiles (see Fig.~\ref{fig4}(d), bottom) at a close distance of $92\mu$m, which matches the theoretical resolution limit of the MMF-based QPI, $0.5\lambda$/NA. The results are presented in Fig.~\ref{fig3}(d). 
The cross-sections along the peaks shown in Fig.~\ref{fig4}(e) demonstrate the excellent agreement between the original phase sample and the reconstructed image, confirming the resolution enhancement.

To summarize, we propose and experimentally demonstrate MMF QPI based on random complex speckle illumination. We have developed the necessary mathematical framework, conducted simulations, and completed experimental validation. The proposed approach can be easily generalized to arbitrary complex patterns and samples and further utilized in various imaging configurations, including transmission or reflection modes. The simulations and experimental validation reveal about $1.5$ times resolution enhancement compared to conventional coherent imaging.
The spatial resolution and imaging speed can be further improved by advanced computational approaches, such as computational machine learning \cite{li2022generative, li2023super} and compressive sensing~\cite{wu_imaging_2021}.
In our experiments, the imaging speed was limited by the camera frame rate. However, the proposed approach does not require a camera sensor and relies solely on a center of mass position, which can be measured by a high-speed PSD. Therefore the imaging speed is fundamentally limited by the scanning rate and the total number of measurements. A digital micromirror device with a maximum speed of $22$~kHz and $400$ measurements leads to a video-rate QPI of $50$ frames per second.
The proposed MMF-based noniterative QPI paves the way toward high-speed ultracompact phase sensors with many applications in biology, medicine, and the semiconductor industry.

\backmatter

\bmhead{Supplementary information}

\bmhead{Acknowledgments} 
This work has been partially carried out within ARCNL, a public-private partnership between UvA, VU, NWO and ASML, and was financed by ‘Toeslag voor Topconsortia voor Kennis en Innovatie (TKI)’ from the Dutch Ministry of Economic Affairs and Climate Policy. We thank Mara Niamh Cerise van der Meulen (University of Amsterdam) for her assistance in derivations of Fourier transform theorems and fruitful discussions. We thank Marco Seynen (AMOLF) for his help in programming the data acquisition software. We thank Dr. Stefan Lehmann (ARCNL) for his assistance in the construction of the experimental setup.
The design of the color maps was carried out using the MATLAB package ColorBrewer~\cite{Stephen23}.
\section*{Declarations}

\noindent
The authors declare no conflicts of interest

\bibliography{sn-bibliography}

\end{document}